# Metastable Cation-Disordered Niobium Tungsten Oxides as Li-ion Battery Anode Materials

Basirat Raji-Adefila[†], You Wang[†], Alexandra Outka[†], Hailey Gonzales[†], Kory Engelstad[‡], Sami Sainio[§], Dennis Nordlund[§], Shan Zhou*[,‡], Dongchang Chen*[,†]

[†] *Department of Chemistry and Chemical Biology, University of New Mexico, Albuquerque, New Mexico 87131, USA*

[‡] *Nanoscience & Biomedical Engineering, South Dakota School of Mines & Technology, Rapid City, South Dakota, 57701*

[§] *Stanford Synchrotron Radiation Lightsource, SLAC National Accelerator Laboratory, Menlo Park, CA, 94025 USA*

*\* Corresponding author's email: <u>chend@unm.edu</u>, <u>shan.zhou@sdsmt.edu</u>*

**Abstract.** Metastable cation-disordered compounds have greatly expanded the synthesizable compositions of solid-state materials and drawn sharp attention among battery electrochemists. While such a strategy has been very successful in a few well-known structures, such as rock salts, metastable cation-disordered materials for other structural types, especially for non-close packed structures, are peculiarly underexplored. In this work, we develop a new series of fully cation-disordered metastable niobium tungsten oxides (NWOs) with a simple structure, and name this new structural type "anti-Li$_3$N". Furthermore, we find that metastable anti-Li$_3$N NWOs transform to a cation-disordered cubic structure when applied as a Li-ion battery anode, highlighting an intriguing non-close packed – close packed conversion between two cation-disordered phases, as evidenced in various physicochemical characterizations, in terms of diffraction, electronic, and vibrational structures. This work enriches the structural and compositional space of NWO families, cation-disordered solid-state materials, and the working mechanisms of Li-ion battery anodes.

Cation disordered ionic solids, in which all cations are randomly dispersed in an anionic sublattice and crystallographically equivalent, has greatly expanded the synthesizable chemical space of solid-state materials and led to new discoveries in materials functionalities, such as in Li-ion batteries (LIBs). [1, 2, 3] While structural families of solid-state materials are extremely rich, cation disordering has mostly been realized in close packed anionic sublattices, such as the well-known face centered cubic (fcc) type and hexagonal close packed (hcp) type. Introducing cation disordering in a close packed anionic sublattice generally leads to simple structures that are analogous to well-known structural types. For example, randomly dispersing octahedrally coordinated cations in a fcc oxygen lattice generates cation-disordered rock salts,[3] isostructural with NiO (Fig. 1a). Non-close packed structures are another extensive materials category and may present fascinating properties.[4, 5, 6, 7] However, different to close packed structures, realizing cation disordering in non-close packed structures is much less explored and has great potentials in creating simple, yet entirely novel structural types to achieve intriguing properties and applications. Therefore, such insufficient research efforts have resulted in a large variety of solid-state materials underexplored, and fundamentally obstructed the exploration of desired materials functionalities that could be derived from this uncharted field.

The class of niobium tungsten oxides (NWOs) is a series of materials that provide rich opportunities for exploration of the possibilities of cation disordering in non-close packed structures. In the recent decade, traditional types of NWOs have found their applications as LIB anode materials and have made a remarkable impact on the research community.[8, 9, 10] Featuring various topological combinations of $ReO_3$-type motifs (Fig. 1b), these oxides have fascinatingly interesting non-close packed lattices,[11, 12] including distorted $ReO_3$-type structures (e.g. $WO_3$);[13, 14] Wadsley-Roth (WR)-type structures (e.g. $H$-$Nb_2O_5$ and $Nb_{16}W_5O_{55}$, etc.);[15, 16] and Tungsten-Bronze (TB)-type structures (e.g. $Nb_9W_8O_{47}$, $Nb_4W_7O_{31}$, etc.).[17, 18] Whether a simple cation-disordered version of NWO that differs from all these well-understood structures can be achieved is an intriguing question (Fig. 1b).

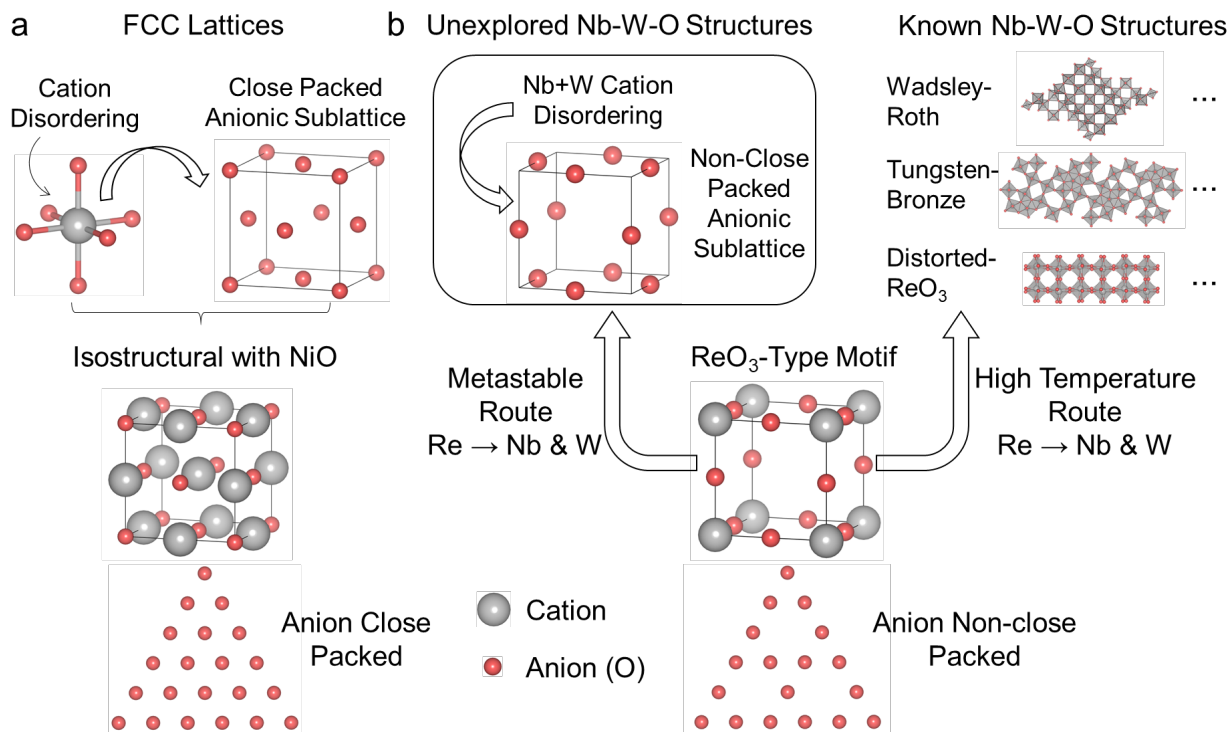

**Fig. 1:** (a) Schematic of octahedrally coordinated cation-disordering in a fcc lattice is isostructural with NiO. (b) Schematic of exploring cation-disordered NWO via a metastable route and known NWO structures obtained from high temperature routes. The ReO$_3$-type motif is the origin of NWO structures. High-temperature NWOs include three major structural varieties, including distorted ReO$_3$, Wadsley-Roth type, and Tungsten-Bronze type. In both a and b, the close packed anions of NiO and non-close packed anions of ReO$_3$ are also shown.

## Results

### Synthesis and Characterizations of Metastable NWOs

In this work, we develop a new series of fully cation-disordered metastable niobium tungsten oxides (NWOs). We found that three Nb/W ratios, including 2:1, 4:1, and 1:0, lead to a very similar type of simple X-ray diffraction (XRD) pattern of the final products with a nanocrystalline-like morphology (Supplementary Fig. 1). The three compositions are denoted as MS-Nb$_2$WO$_8$, MS-Nb$_4$WO$_{13}$, and MS-Nb$_2$O$_5$ (MS: metastable). The simple XRD patterns of MS NWOs are drastically different to those of the parent structures and those of the common structural types of NWOs, as shown in Supplementary Fig. 2a-b. Interestingly, XRDs of the MS NWOs share considerable similarities to the low-temperature phases (i.e. T phase) of Nb$_2$O$_5$ and Ta$_2$O$_5$, but do not present the superstructural diffraction features of the two compounds (Supplementary Fig. 2c).

To unravel this structural type, we performed a detailed structural analysis, as elaborated below (Fig. 2a). By expanding the ReO$_3$-motif, moving the relative locations of atoms, and decreasing the site occupancy of oxygen, we obtained a 2×1 ReO$_3$-like NWO with an alternating transition metal (TM) arrangement, as the primitive structural feature of T-Nb$_2$O$_5$ and T-Ta$_2$O$_5$. XRD simulation of this primitive structural feature is highly consistent to the XRD of the NWOs and T-Nb$_2$O$_5$ without superstructural diffractions (Supplementary Fig. 3). The structure of the MS NWOs can be obtained from another approach. We use alpha-Li$_3$N as the starting structure and apply an anti-site swap that replaces Li site with O and N site with TM, similar to the anti-site swap between antifluorite (e.g. Li$_2$O) and fluorite (e.g. CaF$_2$). Subsequently, after a change of symmetry properties and decreasing the site occupancy of oxygen on the Nb-W-O plane, the resulting unit cell has a fascinatingly simple monoclinic *P2/m* lattice, corresponding to a new type of anti-Li$_3$N-like non-close packed cation-disordered structure. The term "fully cation-disordered" is supported by the fact that all cation sites are crystallographically equivalent in the anti-Li$_3$N unit cell, similar to the full cation disorderness in cation-disordered rock salts (i.e. all Ni sites are equivalent in NiO).[19, 20, 21, 22, 23] The "anti-Li$_3$N" unit cell and the primitive structural feature of T-Nb$_2$O$_5$/Ta$_2$O$_5$ can interconvert via a coordinate transformation (details are described Supplementary Fig. 3).

Based on the monoclinic anti-Li$_3$N structural model, the experimental XRD of the NWOs can be well fitted (Fig. 2b). The cell parameters demonstrate a systematic trend as a function of the composition (Supplementary Table 1). In general, more Nb content leads to an expansion of the a and b axes and the cell volume, which is likely a result of a larger ionic radius of Nb$^{5+}$ than W$^{6+}$.[24] Additionally, we applied Monte Carlo method to evaluate the role of site randomness (i.e. coordinates and occupancy) in XRD pattern, as described in Supplementary Note 1. It indicates that the anti-Li$_3$N structure can be considered as a general structural type where TMs of the a-c plane of are alternatingly spaced (in contrast to the evenly spaced TM in ReO$_3$), oxygens on the a-c plane are disorderly dispersed, and TMs on the a-c planes are connected via cornering oxygen along b axis, as shown in Supplementary Fig. 4. The metastable nature of the anti-Li$_3$N NWOs is demonstrated in Supplementary Fig. 5. After a high-temperature annealing process, all anti-Li$_3$N NWOs fall into the most stable forms of the high temperature phase diagram, as discussed in Supplementary Note 2.

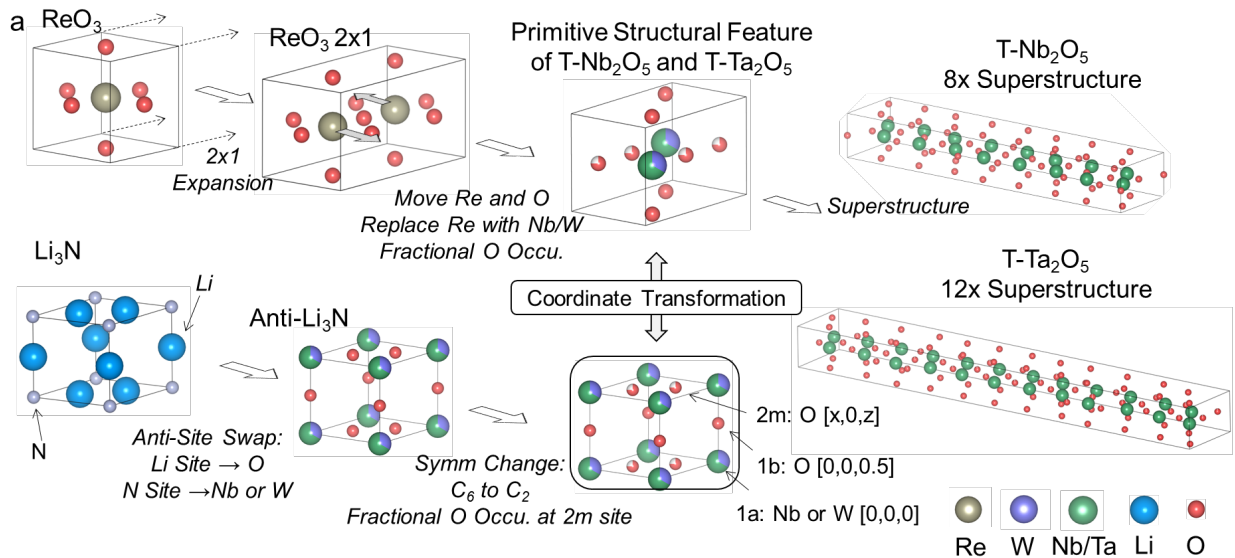

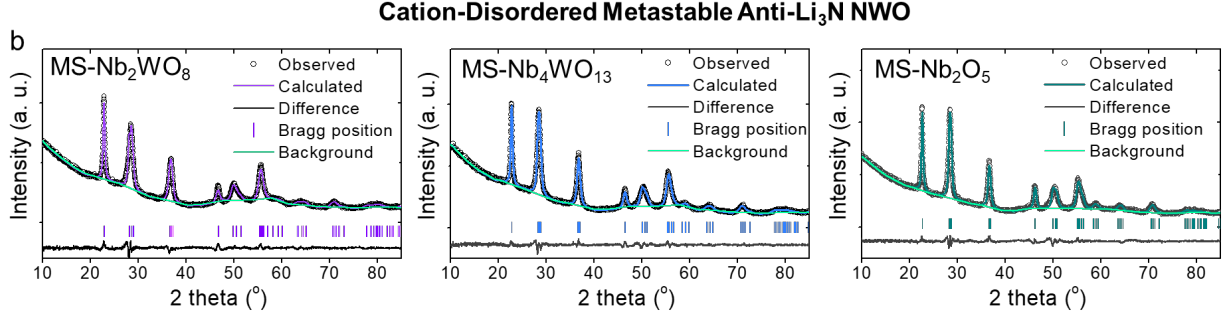

**Fig. 2:** (a) Schematic of obtaining the primitive structural feature of T-$Nb_2O_5$ and T-$Ta_2O_5$ by manipulating the structure of $ReO_3$; superstructural expansion of the primitive structural feature to T-$Nb_2O_5$ and T-$Ta_2O_5$; converting hexagonal $Li_3N$ to monoclinic anti-$Li_3N$ NWO; and the relationship between the primitive structural feature of T-$Nb_2O_5$/$Ta_2O_5$ and the monoclinic anti-$Li_3N$ NWO. (b) Le Bail fittings of the XRDs of MS-$Nb_2WO_8$, MS-$Nb_4WO_{13}$, and MS-$Nb_2O_5$ based on the anti-$Li_3N$ model.

In parallel to the XRD analyses, we applied transmission electron microscopy (TEM) and selected area electron diffraction (SAED) analysis to MS-$Nb_2WO_8$. The low-magnification TEM image (Fig. 3a) demonstrates the nanocrystalline morphology of the particle. An SAED pattern of MS-$Nb_2WO_8$ is shown in Supplementary Fig. 6 and matches the experimental XRD of MS-$Nb_2WO_8$. Fig. 3b and c show two high magnification TEM images. Based on Fast Fourier Transform (FFT) analysis shown in the inset, the spatial frequency of the fringe shown in Fig. 3b is ~ 2.6/nm, corresponding to (010) diffraction of anti-$Li_3N$ MS-$Nb_2WO_8$. Fig. 3c highlights a set of interwoven-like lattice fringes. Based on the spatial frequencies of the two diffractions, we assign them to (001) and (100) diffractions of the anti-$Li_3N$ lattice. The clearer lattice fringes shown in Fig. 3b and c were obtained by inverse FFT (IFFT) treatment of the corresponding diffraction spots.

The enlarged image of the IFFT treatment of (001) and (100) diffractions matches well with the anti-Li$_3$N unit cell projected on the a-c plane. Similarly, the packing of Nb and W along the unique axis (b axis) matches the (010) fringe, further corroborating the structure of anti-Li$_3$N MS NWO.

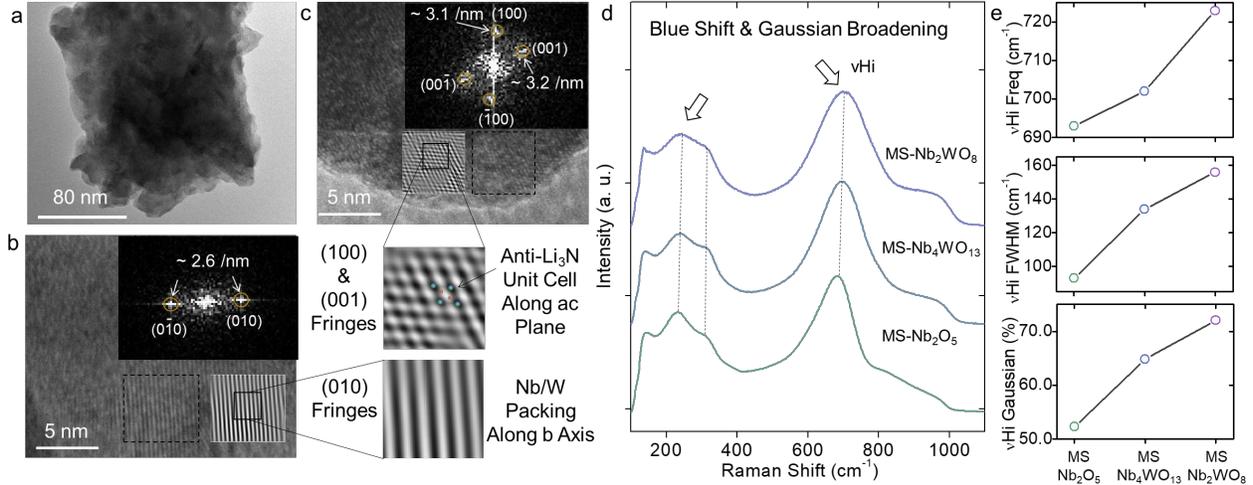

**Fig. 3:** (a) Low-magnification TEM image of MS-Nb$_2$WO$_8$. (b) and (c) Two high-magnification TEM images of MS-Nb$_2$WO$_8$. The insets show the FFT patterns of the regions enclosed in the dash line boxes. A high-percentile filter is applied to the FFT patterns to highlight the marked diffraction spots. The IFFT images of the marked diffraction spots are shown beside the regions enclosed in the dash line boxes. 2 nm x 2 nm magnified images of the IFFT are also shown. (d) Comparison of Raman spectra of the MS NWOs, including MS-Nb$_2$WO$_8$, MS-Nb$_4$WO$_{13}$, and MS-Nb$_2$O$_5$. (e) Quantified band frequency, band full-width at half maximum (FWHM), and Gaussian component for the high wavenumber band centered at 690-730 cm$^{-1}$ ($v$Hi) as a function of the composition of MS NWOs.

Raman spectroscopy is used to analyze the structure of the metastable NWOs from a perspective of lattice vibration. As shown in Fig. 3d, the Raman spectra of all the NWOs demonstrate a set of Raman modes very similar to the Raman bands of T-Nb$_2$O$_5$.[5, 25] Such consistency originates from the fact that the anti-Li$_3$N NWOs have the primitive structural features of T-Nb$_2$O$_5$. A moderate blue shift is observed as the W content increases (detailed fitting is shown in Supplementary Fig. 7), which is a clear indication of mode-hardening due to the higher positive charge of W$^{6+}$ and lattice size contraction of the anti-Li$_3$N unit cell with a higher W content. Moreover, while the Raman bands of MS-Nb$_2$O$_5$ are more of Lorentzian characters, Gaussian broadening is found more significant for NWOs with a higher W content. It indicates that a higher degree of Nb-W mixing in the anti-Li$_3$N lattice causes more Gaussian-like randomness in vibration mode frequencies of metal-oxygen bonds.

**Electrochemical Behavior**

The electrochemical behavior of the metastable NWOs as LIB anode materials was systematically investigated. Voltage profiles of a few selected cycles of MS-Nb$_2$WO$_8$ are shown in Fig. 4a. The 1$^{st}$-cycle voltage profile suffers from considerable amount of irreversible capacity loss during the initial lithiation process, which is commonly found for LIB anodes, due to reductive decomposition of the electrolyte solvent.[26, 27, 28, 29] The 2$^{nd}$ cycle differential capacity (dQ/dV) profiles of the MS NWOs are shown in Fig. 4b. A higher W content leads to more contribution in the low voltage region with a flat profile. A higher Nb content leads to more contribution of the higher-voltage redox peaks (Fig. 4b). The dQ/dV profiles of other cycles demonstrate a similar trend (Supplementary Fig. 9). The rate capability and cycling stability of the MS NWOs are shown in Fig. 4c and d. The 1$^{st}$-cycle specific delithiation capacities for MS-Nb$_2$WO$_8$, MS-Nb$_4$WO$_{13}$, and MS-Nb$_2$O$_5$ are 488, 442, and 415 mAh/g respectively, while the values for the 200$^{th}$ cycle are 200, 205, and 211 mAh/g, respectively. Electrochemical cycling with a higher low-voltage cutoff (0.5 V) leads to lower initial capacities but a better capacity retention, with the 1$^{st}$-cycle specific delithiation capacities of 299 (MS-Nb$_2$WO$_8$), 288 (MS-Nb$_4$WO$_{13}$), and 274 (MS-Nb$_2$O$_5$) mAh/g and the 200$^{th}$-cycle specific delithiation capacities of 143 (MS-Nb$_2$WO$_8$), 159 (MS-Nb$_4$WO$_{13}$), and 162 (MS-Nb$_2$O$_5$) mAh/g. Various-rate cycling shows that MS-Nb$_4$WO$_{13}$ is most capable of high-rate operation (100 mAh/g at 25 C) among the NWOs. These electrochemical data reveal a clear trend that more W is favorable for initial capacity, more Nb is favorable for cycling stability, and the best rate capability can be achieved via a proper combination of Nb and W. Additionally, it's worth noting that the electrochemical performance of MS-Nb$_2$WO$_8$ is significantly greater than that of the thermodynamically stable counterpart, suggesting the opportunity to explore more desirable functionalities from metastable chemical spaces, as discussed in Supplementary Note 5.

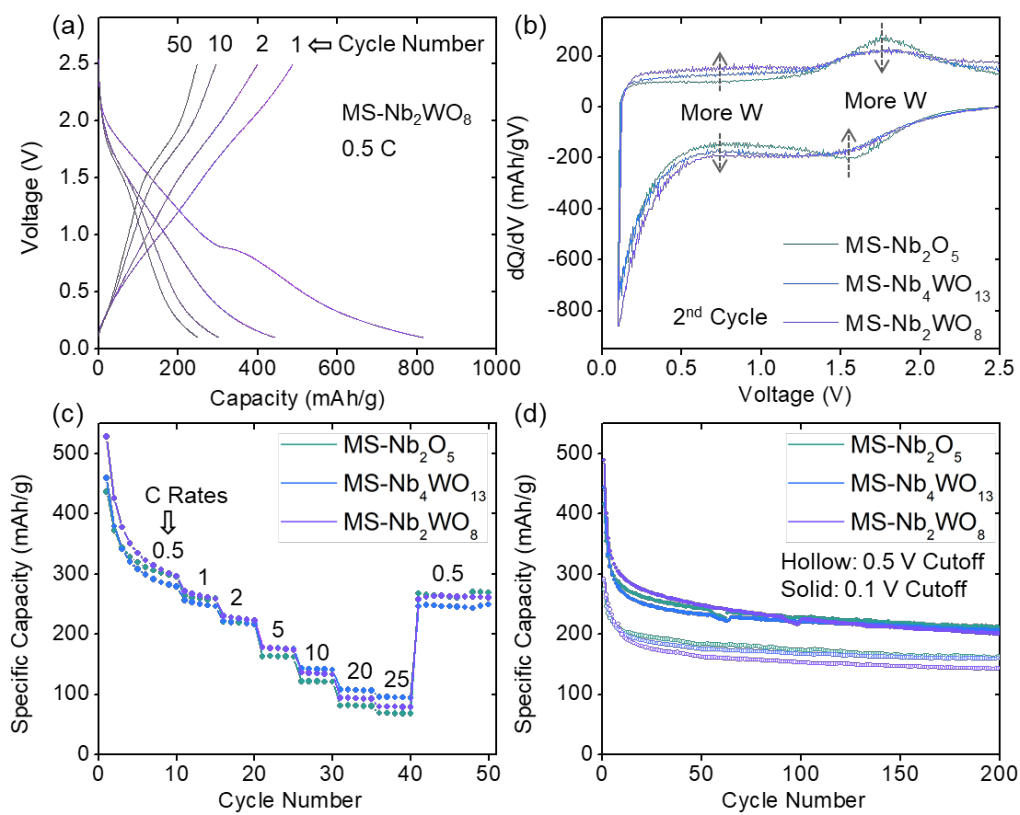

**Fig. 4.** (a) Voltage profiles of MS-Nb$_2$WO$_8$ at 1st, 2nd, 10th, and 50th cycles. The current rate was 0.5 C. 1 C = 200 mAh/g. Similar voltage profiles of other MS NWOs are shown in Supplementary Fig. 8. (b) Differential capacity (dQ/dV) profiles of the MS NWOs derived from the 2nd-cycle voltage profiles. The current rate was 0.5 C. The cathodic contribution when the voltage approaches the lower cutoff (0.1 V) is most likely due to reductive electrolyte decomposition. dQ/dV profiles of other cycles are shown in Supplementary Fig. 9. (c) Specific delithiation capacity retention of various MS NWOs at different C-rates. The low voltage cutoff is 0.1 V. (d) Specific delithiation capacity retention of various MS NWOs over 200 cycles at 0.5 C.

**Electrochemistry-induced Phase Change**

We selected MS-Nb$_2$WO$_8$ as the representative composition for studying the details of electrochemically induced structural change, as shown in Fig. 5a. As MS-Nb$_2$WO$_8$ is electrochemically reduced and incorporates Li, the pristine XRD peaks shift to lower angles (states A-C), an indication of Li intercalation without a phase change. A significant decrease of (010) diffraction peak intensity (~ 2θ ~ 22°) is observed, suggesting the weakening of long-range ordering along the unique b axis. As the voltage further decreases (states C-E), the pristine XRD peaks gradually disappear and a set of new *Fm-3m*-like diffraction peaks emerge. The drastic diffraction pattern change demonstrates that the pristine anti-Li$_3$N monoclinic *P2/m* cation-disordered lattice transforms to an *Fm-3m* cation-disordered lattice, highlighting a critical phase change from a non-close packed oxygen sublattice to an fcc-like close packed one. The cation-disordered nature of the *Fm-3m* lattice is supported by the fact that *Fm-3m* is the space group of fcc oxygen sublattice and any long-range cation ordering will break the *Fm-3m*-like symmetry. At the 2$^{nd}$ fully reduced state (state G), the *Fm-3m*-type diffraction features are more pronounced, indicating a single-phase *Fm-3m* cubic lattice is formed at this state. Evolution of phase structure and lattice parameters obtained from Le Bail fitting (Supplementary Fig. 11) are summarized in Fig. 5b. Li intercalation in pristine MS-$Nb_2WO_8$ gradually lengthens the a, b and c axes (Fig. 5b, A-C). After the critical phase change to a cubic *Fm-3m* lattice between 1-0.5 V, the size of the cubic lattice continues to expand with Li intercalation (D-E). Subsequent delithiation/lithiation results in a reversible contraction/expansion of the cubic *Fm-3m* cation-disordered lattice (states E-G), similar to the energy storage mechanism of cation-disordered rock salts.

TEM analyses were also applied to investigate the crystalline structure of MS-$Nb_2WO_8$ at the fully electrochemically reduced state (i.e. 0.1 V of the 2$^{nd}$ cycle), as shown in Fig. 5c-d. A typical *Fm-3m* diffraction aligned along [01-1] zone axis is observed in the SAED pattern. Each diffraction is composed of two spots, which could be a result of the probed particle consisting of two crystal grains oriented along [01-1]. We also surveyed other particles and observed polycrystalline-like SAED patterns (diffraction rings composed of discrete spots) that are consistent to an *Fm-3m* pattern, as shown in Supplementary Fig. 13. Fig. 5d shows a high-magnification TEM image of the lithiated MS-$Nb_2WO_8$. The observed lattice fringe (~ 4.8/nm, based on FFT analyses) corresponds to the (200) fringe of the *Fm-3m* phase.

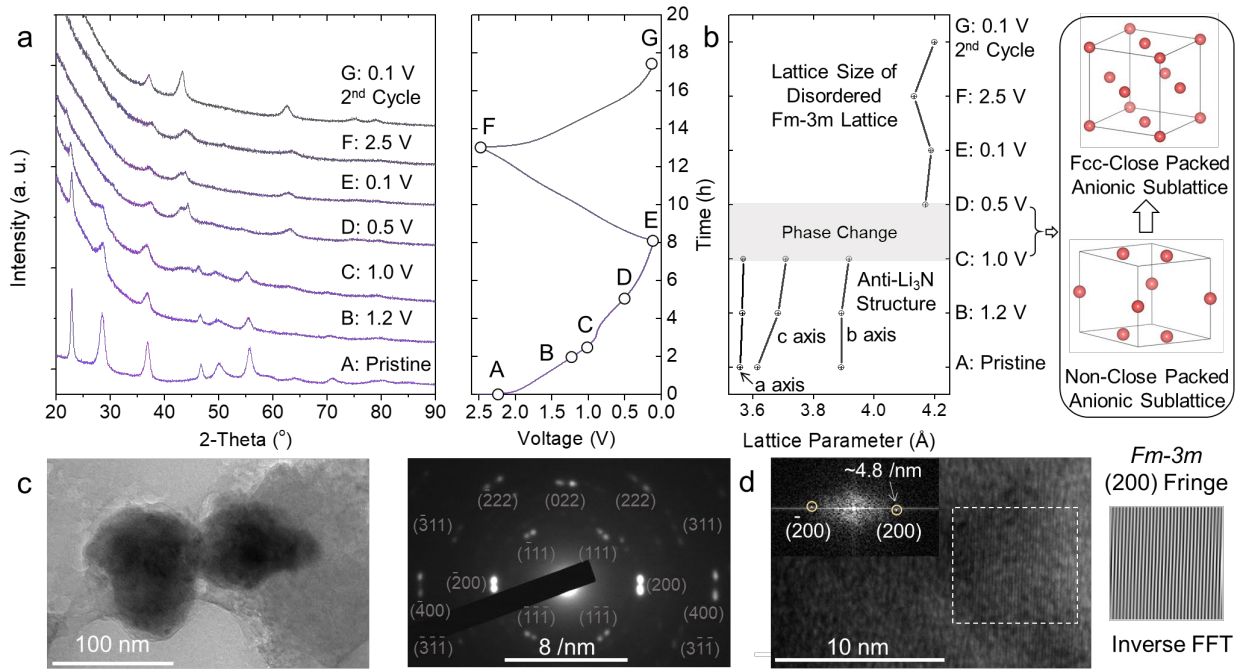

**Fig. 5.** (a) Ex-situ XRD of MS-Nb$_2$WO$_8$ electrodes at various states of charge/discharge (labeled states A-G) and the corresponding voltage profile. The electrodes were peeled off from current collectors. At states D and E, the (200) diffraction of the *Fm-3m* phase is overlapped with a minor peak, which could be a result of (020) diffraction of the residual non-converted anti-Li$_3$N phase. (b) Evolution of lattice constants of the non-close packed anti-Li$_3$N structure, the phase change, and the lattice size of the *Fm-3m* lattice, as a function of states of charge/discharge. (c) Low-magnification TEM image and SAED of an MS-Nb$_2$WO$_8$ electrode discharged at 0.1 V of the 2$^{nd}$ cycle. (d) High-magnification TEM image of an MS-Nb$_2$WO$_8$ electrode discharged at 0.1 V of the 2$^{nd}$ cycle. The inset shows the FFT pattern of the region enclosed in the dash line box. The IFFT image of the two marked diffraction spots is also shown.

Fig. 6a shows the XRDs of the MS NWO electrodes at the 2$^{nd}$ fully reduced state (0.1 V), showing all MS NWOs converted to an *Fm-3m* crystal structure. A higher Nb content also leads to a greater lattice parameter of the *Fm-3m* cell, as shown in Fig. 6b. Raman spectroscopy was also used to probe the electrochemically induced phase transformation from a lattice vibration perspective (Fig. 6c). An exact correspondence is not found between the observed broad band of fully reduced MS NWOs (i.e. at 0.1 V) and Raman spectroscopic features of common TM oxide structures. In a typical *Fm-3m* structure, both 4a and 4b sites (e.g. Na and Cl sites of NaCl, a typical rock salt) are Raman inactive and 8c sites (e.g. Li site of Li$_2$O, a typical antifluorite) involve one Raman-active mode (T$_{2g}$ symmetry).[30] We speculate that the broad overlapped Raman bands are results of vibration modes at 4b site (i.e. oxygen site, general M-O vibration, M=Li, Nb, or W) activated by the break of ideal *Fm-3m* symmetry. The more obvious band at 440 cm$^{-1}$ could be the vibration mode at the 8c site of *Fm-3m* symmetry, as its band position is close to the Li-O mode frequency of various compounds in which Li occupies tetrahedral sites.[31, 32] It should be noted that the more pronounced band (i.e. 440 cm$^{-1}$) is significantly different to $T_{2g}$ of Li$_2$O (~ 520 cm$^{-1}$, Supplementary Fig. 14), in which all 8c sites of an *Fm-3m* lattice are exclusively occupied by Li.

Furthermore, we resorted to oxygen *K* edge X-ray absorption spectroscopy (XAS) to study the change of electronic structures associated with the phase change (Fig. 6d). The major advantage of using oxygen *K* edge XAS in this study is that it can probe the unoccupied Nb$_{4d}$-W$_{5d}$-O$_{2p}$ states as a whole, without relying on the use of edge energy of a specific element. At the pristine state of MS-Nb$_2$WO$_8$, the pre-edge features a broad band at ~ 531 eV with more spectral weight near the edge, which could be a result of the oxygen coordination field posed on Nb and W in the anti-Li$_3$N structure. At 0.1 V, as the MS-Nb$_2$WO$_8$ transforms into an *Fm-3m* structure, the coordination field of Nb/W may be subsequently altered to an octahedral field, which splits the Nb$_{4d}$-W$_{5d}$-O$_{2p}$ states to a $t_{2g}$ branch and an $e_g$ branch.[33] Considering to the large $t_{2g}$-$e_g$ splitting of 4*d*/5*d* oxides (~ 3.6 eV in general),[33, 34, 35, 36, 37] we consider the sharper doublet peak and the shoulder peak (spaced ~3.5 eV) correspond to the transition from O$_{1s}$ to $t_{2g}$ and $e_g$, respectively. The blue shift of the pre-edge after lithiation is caused by the filling of the Nb$_{4d}$-W$_{5d}$-O$_{2p}$ orbitals due to electrochemical reduction, which lifts the energy of lowest unoccupied states.[33, 37] Similarly, all three NWOs demonstrate similar spectral profiles with comparable edge energies at 0.1 V (Fig. 6e), indicating that TM coordination in all the NWOs change to an octahedral type at the fully lithiated state.

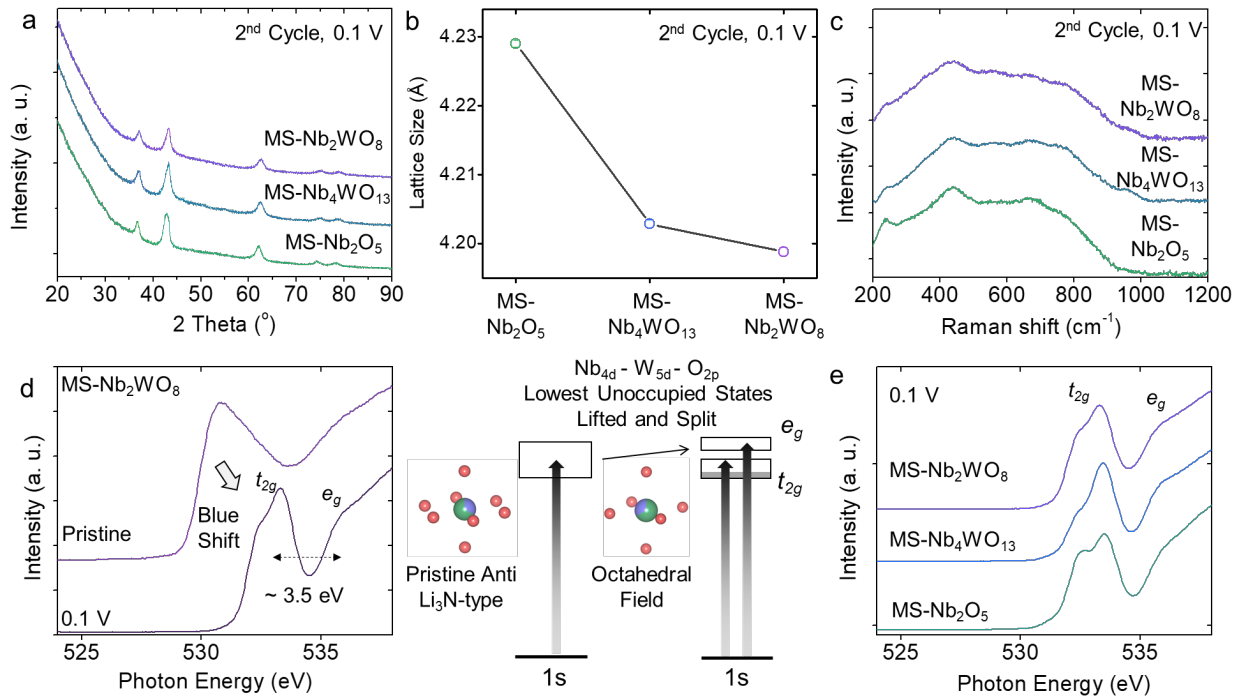

**Fig. 6.** (a) Ex-situ XRDs of various MS NWO electrodes discharged at 0.1 V of the 2$^{nd}$ cycle. (b) The lattice size of the *Fm-3m* lattice shown in (a), as a function of MS NWO composition. (c) Raman spectra of various MS NWO electrodes discharged at 0.1 V of the 2nd cycle. (d) Oxygen *K* edge XAS of MS-Nb$_2$WO$_8$ electrodes at the pristine state and discharged at 0.1 V. The reason for the spectral feature change, the lift and split of the Nb$_{4d}$-W$_{5d}$-O$_{2p}$ lowest unoccupied states, as a result of the electrochemical reduction and phase change is schematically shown. The doublet profile of the $t_{2g}$ may arise from the different transition energies to different spin states due to Coulomb U, which is a common reason for complex O *K* edge spectral features of TM oxides with simple octahedral symmetry.[33, 38] (e) Oxygen *K* edge XAS of various MS NWOs discharged at 0.1 V.

**Discussion**

The structural space of high-temperature NWOs is comprehensively summarized in Krumeich's work.[11] The significant structural difference of our anti-Li$_3$N NWOs to the high-temperature counterparts suggests that a metastable chemical space of NWO exists in parallel to the high-temperature one. To explore other structural possibilities, we also extended the synthesis method to other Nb/W mole ratios and found that the result does not necessary lead to the anti-Li$_3$N-type metastable structures (Supplementary Fig. 15). Metastable NWOs that do not adopt an anti-Li$_3$N-like structure will be reported in future communications.

Among all recently reported phase conversion mechanisms for LIB anode materials, the electrochemically induced phase change to a cubic *Fm-3m* structure observed for various $d^0$ TM oxides, including $V_2O_5$,[39] $Li_2MoO_4$,[40] and amorphous $Nb_2O_5$,[26] has received a great amount of attention. The phase change of MS NWOs from an anti-$Li_3N$ non-close packed structure to *Fm-3m* observed in this work further suggests the generality of the mechanism. As the pristine structures of these TM oxides are highly different, the common conversion process suggests that fcc-like oxygen sublattice may be a thermodynamically favored structure when lithiation voltage is beyond the voltage limit of intercalation. In this work, metastable NWOs completely convert to a *Fm-3m* lattice after only two cycles. We consider the reason for the fast conversion process is because the anti-$Li_3N$ structure is far away from the equilibrium state due to the metastable nature. When the electrochemical potential reaches the region where an fcc oxygen sublattice is favorable, phase conversion will quickly occur. Fundamental thermodynamics and kinetics of the conversion process to *Fm-3m* structures and relationship of these fundamental properties to various materials factors, such as element, TM-O coordination types, and anion packing types, are especially worth investigating.

**Conclusion**

In this work, we address the knowledge gap on cation-disordered compounds in non-close packed structures by discovering a series of new metastable fully cation-disordered niobium tungsten oxides as LIB anode materials. The structural type is named as "anti-$Li_3N$", as the structure is analogous to a result of cation-anion site swap of $Li_3N$. Moreover, we find that MS NWOs transform to a cation-disordered *Fm-3m* cubic structure at lower cell voltages when applied as a LIB anode, indicating an intriguing phase conversion process from a non-close packed oxygen sublattice to a fcc close packed type. The methodology and strategy of the work can be broadly applicable to other fields of solid-state materials, where a "hidden" metastable chemical space may exist in parallel with a more well-known thermodynamically stable counterpart. Application-wise, the study will incite the exploration for clean energy solutions and other key fields of functional materials from all-new structural spaces.

**Acknowledgements.** This work was financially supported by University of New Mexico new faculty startup award, research allocation committee award, and Oak Ridge Associated Universities Ralph E. Powe Junior Faculty Enhancement Award. Use of the Stanford Synchrotron Radiation Lightsource, SLAC National Accelerator Laboratory is supported by the Office of Science, Office of Basic Energy Sciences of the U.S. Department of Energy under Contract No.

DE-AC02-76SF00515. K.E. and S.Z. acknowledge the faculty startup award at the South Dakota School of Mines and Technology.

## Methods

### Characterization and Analyses

X-ray diffraction patterns of the samples were taken with a Rigaku Smart Lab X-ray diffractometer (Cu Kα, 40 kV, 40 mA). A Ni-foil Cu Kβ filter was installed to filter out most of the Kβ diffraction peaks. Recovered cycled electrodes were first thoroughly rinsed with dimethyl carbonate and dried in a glovebox, then pasted on a Kapton tape using a single crystal silicon holder for XRD measurements. All crystal structure visualization, structural manipulation, and XRD simulation were done using VESTA. Le Bail fitting and Rietveld refinements of XRD patterns were performed using the FullProf software. Cell expansion via a Monte-Carlo method was realized via home-written MATLAB codes. Scanning electron microscopy (SEM) images were taken to examine the morphology of the samples via an FEI Helios Nanolab 650 scanning electron microscope. Raman spectra were obtained using a Renishaw inVia Spectro-microscopy system with an Olympus 20 x objective. A He-Ne laser (632.8 nm) was used as excitation of Raman spectra. The polarization ratio of the laser is greater than 500:1. The confocal slit was adjusted to be 10 μm to minimize the spectroscopic contribution of non-confocal signals and to minimize band broadening effects. The monochromator uses an 1800 line/mm grating. A 1024x256 pixel CCD camera was used for spectral acquisition. Raman spectral fitting was performed via a least square method assuming Voigt line profile of each peak. A JEOL JEM-2100 with a LaB$_6$ emitter at 200 kV was used for TEM experiments. Solid samples freshly out of glovebox were immediately dissolved in ethanol and sonicated for 5 min, and then 0.5 μL of the solution was drop-casted onto a TEM grid and left dry for TEM measurements. FFT and IFFT analyses were done via GMS3 software. Oxygen $K$ edge XAS measurements were carried out at BL 10-1 at SSRL. Samples are attached on conductive carbon tapes, which were attached to an aluminum sample holder rod inside an Ar filled glovebox. Spectral measurements were performed on the 31-pole wiggler with a ring current of 350 mA, a 1000 line mm$^{-1}$ grating monochromator with 20 mm entrance and exit slits, and a 1 mm$^2$ beam spot. XAS profiles were collected under ultrahigh vacuum (10$^{-9}$ Torr) using a total electron yield (TEY) detector.

**Electrochemical Testing**

Electrodes fabrication: The composite containing active material and conductive carbon additives was mixed with polyvinylidene fluoride (PVDF) binder (Kynar 2801) in a 9 : 1 weight ratio in an N-methyl-2-pyrrolidone (NMP, Sigma-Aldrich, 99.7%) solvent. The slurry obtained was casted onto copper foil and dried overnight at 100 °C under vacuum. Dried electrodes were cut to disks with an area of 1.6 cm$^2$ and a typical mass loading of 2–3 mg cm$^{-2}$ and were subsequently used for cell testing in 2032-type coin cells. Half-cells were assembled in an argon-filled glovebox with Li foils (Sigma - Aldrich) as counter and reference electrodes and glass microfiber separators. A solution of 1 M LiPF$_6$ in 1 : 1 (v/v) ethylene carbonate (EC) and diethylene carbonate (DEC) (Novolyte Technologies Inc) was used as the electrolyte. The cells were galvanostatically cycled via a Neware Battery Testing System. All electrochemical measurements were carried out at room temperature.